# Europa's Surface Water-ice Crystallinity and Correlations between Lineae and Hydrate Composition

Jodi R. Berdis[1,2], James R. Murphy[1], and Nancy J. Chanover[1]
[1] Astronomy Department, New Mexico State University, Las Cruces, NM 88003, USA; jodi.berdis@jhuapl.edu
[2] Now at: Johns Hopkins University Applied Physics Laboratory, Laurel, MD 20723, USA

## Abstract

Europa's surface composition and evidence for cryovolcanic activity can provide insight into the properties and composition of the subsurface ocean, allowing the evaluation of its potential habitability. One promising avenue for revealing the surface processing and subsurface activity are the relative fractions of crystalline and amorphous water-ice observed on the surface, which are influenced by temperature, charged particle bombardment, vapor deposition, and cryovolcanic activity. The crystallinity observed on Europa's leading hemisphere cannot be reproduced by thermophysical and particle flux modeling alone, indicating that there may be additional processes influencing the surface. We performed a spectral mixture analysis on hyperspectral image cubes from the Galileo Near-Infrared Mapping Spectrometer (NIMS) to identify how surface crystallinity is influenced by physical processing at a high spatial resolution scale. We focus specifically on two image cubes, 15e015 closer to the equator and 17e009 closer to the south pole, both on the leading hemisphere. We performed a nonnegative least-squares spectral mixture analysis to reveal both the non-ice composition and the water-ice crystallinity of the surface. We found that amorphous water-ice dominates the spectrum at the equator and the south pole. We estimated a mean crystallinity of ∼35% within the 15e015 NIMS cube and a mean crystallinity of ∼15% within the 17e009 NIMS cube, which is consistent with ground-based spectroscopically derived crystallinities. We also identified a correlation of magnesium sulfate, magnesium chloride, and hydrated sulfuric acid with lineae and ridges, which may provide evidence for surface processing by upwelling subsurface material.

*Unified Astronomy Thesaurus concepts:* Surface composition (2115); Europa (2189); Spectroscopy (1558)

## 1. Introduction

Surface composition and evidence for cryovolcanic activity on Jupiter's moon Europa can provide insight into the properties and composition of the subsurface ocean based on the potential transport of material between the subsurface ocean and the icy surface. A long-standing debate exists regarding the source of surface materials on Europa; while magnesium and sodium brines may originate from the underlying ocean (Shirley et al. 2010; Brown & Hand 2013; Ligier et al. 2016; Trumbo et al. 2019), sulfuric acid hydrate could be radiolytically produced from interactions between deposited Io sulfur and Jupiter's magnetic field (Kargel 1991; Carlson et al. 2005). Determining the abundances and locations of these materials on the surface can help to reveal the alteration processes occurring both below the surface and at the surface and aid in identifying whether that alteration is a result of endogenic or exogenic processing, or perhaps both.

A combination of hydrated brines, hydrated sulfuric acid, and water-ice is proposed to be present on Europa's surface and likely reproduces features in spectra of Europa's surface (McCord et al. 1998b, 1999; Carlson et al. 1999; McCord et al. 2002; Shirley et al. 2010; Dalton et al. 2012). Most recently, Dalton et al. (2012) found that a mixture of crystalline water-ice, hydrated sulfuric acid ($H_2SO_4$), and sodium and magnesium hydrated salts could best reproduce the spectra in several Galileo Near-Infrared Mapping Spectrometer (NIMS) hyperspectral image cubes.

Previous studies have focused on constraining the water-ice and brine composition of Europa's surface on regional scales in order to identify how surface processes may be influencing different regions of the surface. Most notable in these investigations has been (1) the compositional differences between the trailing and leading hemispheres, where the trailing hemisphere is exposed to considerable surface radiation due to particle bombardment from Jupiter's magnetic field and Io's volcanoes (Carlson et al. 2005); and (2) the compositional differences between "old" terrain (such as ridged plains) and "young" terrain (such as dark smooth plains, which are hypothesized cryovolcanic flow deposits; Greeley et al. 2000; Shirley et al. 2010). Shirley et al. (2010) found that the older ridged plains are dominated by water-ice and the younger dark plains are dominated by hydrated salts. They also identified magnesium sulfate and a higher abundance of sodium-bearing species in the smooth plains, and therefore the presence of these materials could be a result of cryovolcanic flow onto the surface (Shirley et al. 2010).

Dalton et al. (2012) found that the leading hemisphere equatorial plains are dominated by smaller grain sizes (50 and 75 $\mu$m) of crystalline water-ice with low amounts of hydrated salts and sulfuric acid. They also found that the southern high-latitude regions on the leading hemisphere are dominated by slightly larger grain sizes (75 and 100 $\mu$m) of crystalline water-ice with no traceable amounts of hydrated sulfuric acid. On the trailing hemisphere, however, hydrated sulfuric acid dominates, followed by magnesium and sodium sulfate (Dalton et al. 2012). Water-ice grain sizes are dependent on distance from the trailing hemisphere apex, and the distribution of hydrated sulfuric acid is independent of the underlying geological features (Dalton et al. 2012). Ligier et al. (2016) confirmed that hydrated sulfuric acid is correlated with distance from the trailing hemisphere apex (Carlson et al. 2005), and magnesium-bearing species are correlated with geological features, suggesting an endogenic origin.

The relative fractions of amorphous and crystalline water-ice ("crystallinity") are influenced by (a) the thermal relaxation of amorphous into crystalline water-ice, (b) the conversion of crystalline to amorphous water-ice due to charged particle bombardment from interactions with Jupiter's magnetic field, (c) vapor deposition of water-ice as outbursted plume material (in either the crystalline or amorphous form) falls back onto the surface, and (d) any additional cryovolcanic activity such as diapirs that could warm the ice and convert amorphous into crystalline water-ice. Crystalline water-ice and amorphous water-ice are spectrally distinct, and the relative quantities of each may be approximated based on the analysis of certain spectral absorption features. Water-ice absorption features at 1.5, 1.65, and 2.0 $\mu$m increase in depth and shift to longer wavelengths as the fraction of crystalline water-ice increases (Schmitt et al. 1998; Mastrapa et al. 2008).

The crystallinity of Europa's leading hemisphere cannot be reproduced by thermophysical and charged particle bombardment modeling alone (Berdis et al. 2020). This discrepancy may indicate that additional processes have existed in the past (or are currently active) that may be influencing Europa's surface and altering its crystallinity. In this study we aim to identify the spatial distribution of the crystallinity for two locations near the equator and south pole on the leading hemisphere. A higher abundance of crystalline water-ice is expected in regions closer to the equator (compared to regions closer to the poles) owing to the thermal relaxation of water-ice, where the relaxation timescale, as well as the subsequent conversion of amorphous into crystalline water-ice, is dependent on the temperature of the ice (Kouchi et al. 1994; Jenniskens et al. 1998; Mastrapa et al. 2013). For example, amorphous water-ice at $T \sim 115$ K will relax fully into the crystalline phase after a few years, whereas amorphous water-ice at $T \sim 90$ K relaxes into the crystalline phase after $10^5$ yr (Mastrapa et al. 2013). Therefore, a higher relative abundance of crystalline water-ice is expected at equatorial regions, whereas a higher relative abundance of amorphous water-ice is expected at polar regions. Additionally, sublimation of ice and/or migration of plume material may occur at higher latitudes, supplying even more amorphous water-ice to the polar regions (Kouchi et al. 1994; Hansen & McCord 2004).

The presence of the sharp, narrow, 1.65 $\mu$m band, attributed to low-temperature crystalline water-ice (Grundy & Schmitt 1998), has been identified in both space- and ground-based observations of Europa (e.g., Grundy et al. 1999; Hansen & McCord 2004; Grundy et al. 2007). The top $\sim$1 mm of surface water-ice is likely in the amorphous form owing to charged particle bombardment, whereas the water-ice at depths below $\sim$1 mm is likely in the crystalline form since it is protected from radiation by the overlying surface layer (Hansen & McCord 2004). The distribution of amorphous water-ice on Europa's surface is correlated with the intensity of Io's plasma bombardment (i.e., it increases with proximity to the trailing hemisphere apex), whereas crystalline water-ice is more correlated with geomorphological units on the surface, which might trace cryovolcanic activity (Ligier et al. 2016). Ligier et al. (2016) found that the ratio of amorphous to crystalline water-ice is 0.57 globally, indicating a crystallinity of 64%. Berdis et al. (2020) calculated a lower leading hemisphere crystallinity of $\sim$30%. Crystallinity analyses of localized regions at a relatively high spatial resolution, such as specific craters, lineae, chaos terrain, or swaths of plains on Europa, have yet to be published, but they have been successful for other icy satellites such as Saturn's moon Rhea (Dalle Ore et al. 2015).

The compositional interpretation of surface material using spectroscopy can be achieved through spectral mixture theory, the concept that a single observed spectrum is the combination of various pure material spectra, or endmembers, weighted by the respective fractional abundance that exists in that single observation. Spectral mixture analyses have been performed on hyperspectral image data for a majority of the solid planetary bodies in the solar system, including the Moon (Li & Mustard 2003), Mars (Combe et al. 2008; Liu et al. 2016), asteroids (Combe et al. 2015), Saturnian satellites such as Titan (Le Corre et al. 2009; McCord et al. 2008), and Jovian satellites such as Europa (Shirley et al. 2010; Dalton et al. 2012).

Hapke scattering theory provides a method for estimating the bidirectional reflectance of a material by taking into consideration the scattering effects, viewing geometry, and real and imaginary indices of refraction for that material (Hapke 1981). When optical constant information has been produced for a material, it is possible to synthesize a reflectance spectrum of that material's presence on an airless solid surface given the regolith properties (e.g., porosity, grain size), temperature, illumination, and incident and emission angles of the synthesized observation. Hapke modeling has been frequently used in the literature to aid in performing spectral mixture analyses on airless solid surfaces in order to more accurately synthesize endmember spectra (Lucey 1998; Mustard & Pieters 1989; Hamilton et al. 2005; Warell & Davidsson 2010; Ciarniello et al. 2011; Dalton et al. 2012). When optical constant information does not exist for a given material, cryogenic laboratory reference spectra acquired at environmental conditions similar to those in which the observed materials exist may alternatively be used; however, properties of the cryogenic laboratory materials such as viewing geometry and grain size are less likely to match those of the observed materials.

This investigation uses a spectral mixture approach to identify the abundances of pure materials, or endmembers (including amorphous and crystalline water-ice, hydrated sulfuric acid, and brines), that exist across Europa's surface as observed by the Galileo NIMS instrument. We built on previous work by Dalton et al. (2012) by (1) introducing amorphous water-ice into the spectral mixture analysis process alongside crystalline water-ice, (2) using updated optical constants for amorphous and crystalline water-ice (Mastrapa et al. 2008), (3) including a more diverse cryogenic reference spectral library (Hanley et al. 2014), and (4) analyzing the spatial variation of endmember abundances across each NIMS observation, rather than averaging multiple pixels together to represent a specific type of terrain/material. We used hyperspectral image data from the Galileo NIMS instrument (Carlson et al. 1992; Section 2.1), synthetic amorphous and crystalline water-ice spectra at multiple grain sizes produced by Hapke theory (Hapke 1981, 1984, 1986; Section 2.2), and cryogenic laboratory reference spectra (Section 2.3). We implemented the spectral mixture analysis technique to identify abundances of endmember materials within each NIMS pixel separately (Section 2.4). Results for the leading hemisphere equatorial and south polar NIMS observations are presented in Sections 3.1 and 3.2, respectively, and we discuss our findings in Section 3.3 and our conclusions in Section 4.

Table 1
Description of the NIMS Observations Used in This Study

| Observation Name | Observation Date | Latitude | Longitude | Incident Angle | Emission Angle | Phase Angle |
| --- | --- | --- | --- | --- | --- | --- |
| 15e015 | 31 May 1998 | 7.3° N | 114° W | 44°.7 | 41°.8 | 86°.4 |
| 17e009 | 26 Sep 1998 | 63° S | 120° W | 68°.2 | 51°.7 | 69°.4 |

## 2. Data and Methodology

We used hyperspectral image data from the Galileo NIMS instrument, synthetic amorphous and crystalline water-ice spectra produced by optical constants from Mastrapa et al. (2008) and Hapke theory for multiple grain sizes, and cryogenic laboratory reference spectra from Dalton et al. (2005), Dalton (2007), Carlson et al. (2005), and Hanley et al. (2014). We implemented the spectral mixture analysis technique to identify abundances of endmember materials within each NIMS pixel separately.

### 2.1. Galileo NIMS Observations

The NIMS instrument on board the Galileo spacecraft acquired spectral image cubes of Europa's surface over the wavelength range 0.7–5.3 $\mu$m at spatial resolutions down to <2 km (Carlson et al. 1992; McCord et al. 1999). Several times throughout the Galileo mission the NIMS instrument operated in "Long Spectrometer Mode," during which ground tracks with a higher spatial resolution were obtained (Carlson et al. 1992). We focused specifically on two image cubes acquired during this mode of operation: 15e015 (7.3° N, 114° W, 3.0 km pixel$^{-1}$) closer to the equator, and 17e009 (63° S, 120° W, 1.5 km pixel$^{-1}$) closer to the south pole, both on the leading hemisphere (Table 1).

We employed the USGS Integrated Software for Imagers and Spectrometers (ISIS3) software to project the image cubes using the latitude and longitude backplanes. Since the map projections that the NIMS data are stored in are not compatible with ISIS3, we used the nocam2map ISIS3 application to map the NIMS data using the longitude and latitude backplanes (https://astrogeology.usgs.gov/search/map/Docs/PDS/cahill_galileonims_pdart/cahill_galileonims_pdart). The wavelength regions that had duplicate measurements due to overlapping detectors were averaged at the same average spectral resolution (0.0153 $\mu$m for the 15e015 NIMS cube and 0.015 4 $\mu$m for the 17e009 NIMS cube) for our wavelength range of interest, 1.25–2.45 $\mu$m.

### 2.2. Hapke Reflectance Modeling

We employed the Hapke surface-scattering radiative transfer model (Hapke 1981, 1984, 1986) as outlined in Grundy (1995) to produce synthetic spectra of Europa's surface water-ice as a function of viewing geometry and surface material scattering properties. We used optical constants, i.e., the real and imaginary parts of the index of refraction ($n$ and $k$, respectively), for amorphous and crystalline water-ice from Mastrapa et al. (2008). Dalton et al. (2012) produced synthetic crystalline water-ice spectra for the spectral analysis of several NIMS cubes; however, not only did we use more recently published sets of optical constants, but we also included amorphous water-ice along with crystalline water-ice in our study (Figure 1).

The overall spectral reflectance and depths of absorption features of crystalline and amorphous water-ice are dependent on grain size (Clark 1981). Therefore, we produced synthetic spectra of crystalline and amorphous water-ice for grain sizes of 25, 50, 75, 100, 150, 200, and 250 $\mu$m. For the 15e015 NIMS equatorial observation we used water-ice optical constants at $T = 120$ K to represent the approximate temperature near the equator, and for the 17e009 NIMS south polar observation we used water-ice optical constants at $T = 90$ K to represent the approximate temperature near the south pole (Rathbun et al. 2010). The incidence, emission, and phase angles for each observation are provided in Table 1. We assumed a grain porosity of 0.1 (Johnson et al. 2017) to compute the compaction parameter (Hapke 1963; Buratti 1985), an asymmetry factor of −0.15 (Buratti 1983, 1985), and a mean macroscopic roughness parameter of 10° (Domingue et al. 1991). Figure 2 displays an example subset of synthetic reflectance spectra calculated for a range of grain sizes for the Hapke-modeled amorphous and crystalline water-ice.

### 2.3. Laboratory Reference Spectra

Whereas Mastrapa et al. (2008) produced recent optical constants for amorphous and crystalline water-ice, optical constants of our reference brine materials for the specific temperatures and pressures that exist on Europa's surface are not currently available in the literature. Therefore, we used reflectance spectra from Dalton et al. (2005), Dalton (2007), Carlson et al. (2005), and Hanley et al. (2014) for our brine materials. Shirley et al. (2010) and Dalton et al. (2012) included spectra of sulfuric acid hydrate ($H_2SO_4 \cdot 8H_2O$), epsomite ($MgSO_4 \cdot 7H_2O$), hexahydrite ($MgSO_4 \cdot 6H_2O$), mirabilite ($Na_2SO_4 \cdot 10H_2O$), and bloedite ($Na_2Mg[SO_4]_2 \cdot 4H_2O$) in their cryogenic reference library; however, they did not find detectable amounts of $MgSO_4 \cdot 7H_2O$ or $Na_2Mg[SO_4]_2 \cdot 4H_2O$. Ligier et al. (2016) included the above reference materials in addition to several magnesium and sodium brines from Hanley et al. (2014) in their reference library. Our reference library contained the endmembers that were found in detectable quantities in the 15e015 and 17e009 NIMS cubes by Dalton et al. (2012) and all of the laboratory-produced brines by Hanley et al. (2014), which included a variety of magnesium and sodium hydrated salts and potassium chloride (KCl).

Reference spectra for the endmembers used by Dalton et al. (2012) were acquired at either 100 K or 120 K, and the endmembers published in Hanley et al. (2014) were acquired at 80 K. Since the grain sizes and temperatures of the reference endmembers are fixed, it should be noted that differences in the reference endmember spectrum and the spectrum of that substance as it exists on Europa's surface in these NIMS observations may differ and can add some uncertainty to the resulting model fits. Furthermore, as mentioned previously, changes to the reference cryogenic spectra as a result of viewing geometry have also not been included. Additional laboratory work on hydrated brines at different temperatures, grain sizes, and viewing geometries is needed in order to more accurately represent the material that exists on Europa's surface. We also included a spectrally flat neutral dark component to account for possible carbonaceous species (McCord et al. 1998a), differences in lighting between Europa's surface and the laboratory reference material, and/or albedo variations due to illumination or topography across the image cubes (Cheek & Sunshine 2020). We defined this dark absorber endmember as a constant reflectance of 0.001 at all wavelengths.

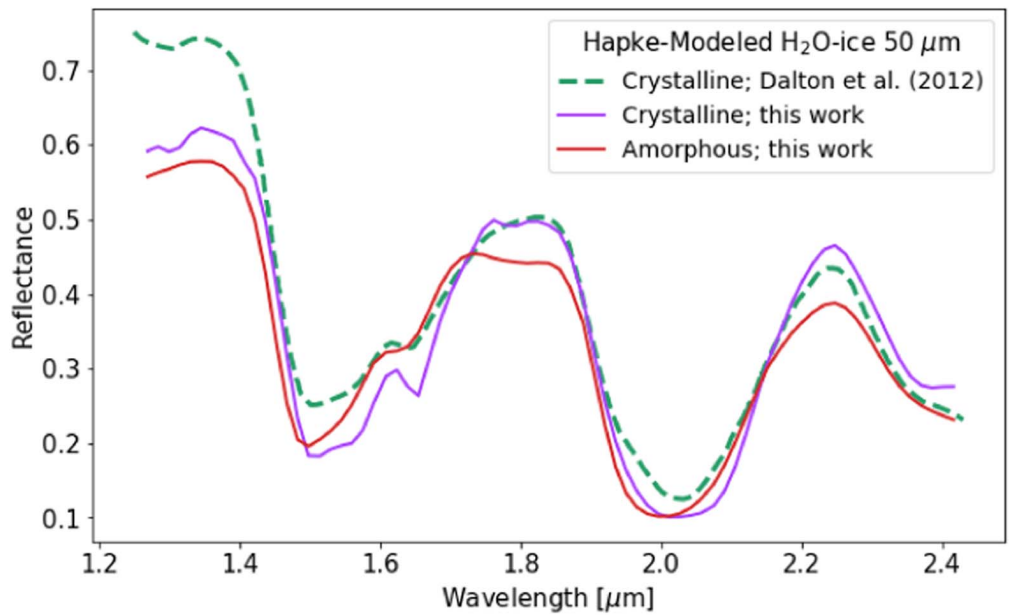

**Figure 1.** Synthetic water-ice spectra with grain sizes of 50 μm produced for a viewing geometry that matches the location of the 15e015 NIMS cube and resampled at a spectral resolution that matches the 15e015 NIMS cube. We used a more recent set of optical constants, leading to slight differences in the shape of our water-ice absorption features compared to those used by Dalton et al. (2012). We included amorphous and crystalline water-ice spectra in our study, whereas Dalton et al. (2012) only included crystalline water-ice.

Figure 3 displays the cryogenic reference library, and Table 2 shows a list of our spectral reference library, including all cryogenic materials and synthetic water-ice from Section 2.2. All spectra were resampled to the spectral resolution of the NIMS data sets.

The wavelength range of 1.25–2.45 μm was specifically selected to maximize spectral overlap between the optical constants, reference materials, and NIMS instrument. Hanley et al. (2014) published results of their laboratory studies in the wavelength range 0.35–2.5 μm, Dalton et al. (2005) published results from 0.7 to 2.45 μm, and Dalton (2007) and Carlson et al. (2005) published results from 1 to 2.5 μm. An instrumental band gap exists in the NIMS data from 1 to 1.25 μm. In order to maximize the number of endmembers that we were able to use, we chose to reduce the wavelength range to 1.25–2.45 μm in order to have a more diverse spectral library. We recognize that several notable water-ice features, such as the 3.1 μm Fresnel peak, that serve as particularly useful metrics for distinguishing between amorphous and crystalline water-ice are not included in this study. We do not assume that the abundances calculated in this study are exact values, but rather represent the trend of abundances at these locations and the approximate relative ratios of material abundances as compared against each other.

### 2.4. Spectral Mixture Analysis

The compositional interpretation of surface material using spectroscopy can be achieved through spectral mixture theory (Gross & Schott 1996; Manolakis et al. 2016; and references therein), the concept that a single observed spectrum ($T(\lambda)$), such as a single spatial pixel within a hyperspectral image cube, is the combination of various $N$ pure material spectra ($R_i(\lambda)$), or endmembers, weighted by their respective fractional abundances ($\omega_i$) that exist in that single observation (Manolakis et al. 2016), provided in functional form as

$$T(\lambda) = \sum_{i=1}^{N} \omega_i R_i(\lambda). \quad (1)$$

For example, if a remotely sensed hyperspectral pixel taken of a suburban region on Earth is spatially made of 50% grass and 50% concrete based on fractional surface area, the observed spectrum for that pixel would consist of the sum of $0.5\times$ the reflectance of a pure grass spectrum and $0.5\times$ the reflectance of a pure concrete spectrum at each wavelength. This technique requires some prior knowledge of the endmembers that are likely to exist in that pixel. The most common form of spectral mixture analyses is the linear spectral mixture approach, which assumes that the endmembers are not intimately mixed on the surface, and each photon does not scatter so much that it interacts with more than one endmember, allowing for a linear sum of the contributing weighted endmembers (Dobigeon et al. 2016). Summed combinations of weights for each endmember are then tested against the observations in some form of "goodness-of-fit" test. This study assumes a simplified linear model over a nonlinear and intimately mixed model since the nature of nonlinearity and intimate mixing is not known in this context, at this scale. Tighter constraints on the particle size distribution on the surface are necessary for nonlinear and intimate mixture modeling.

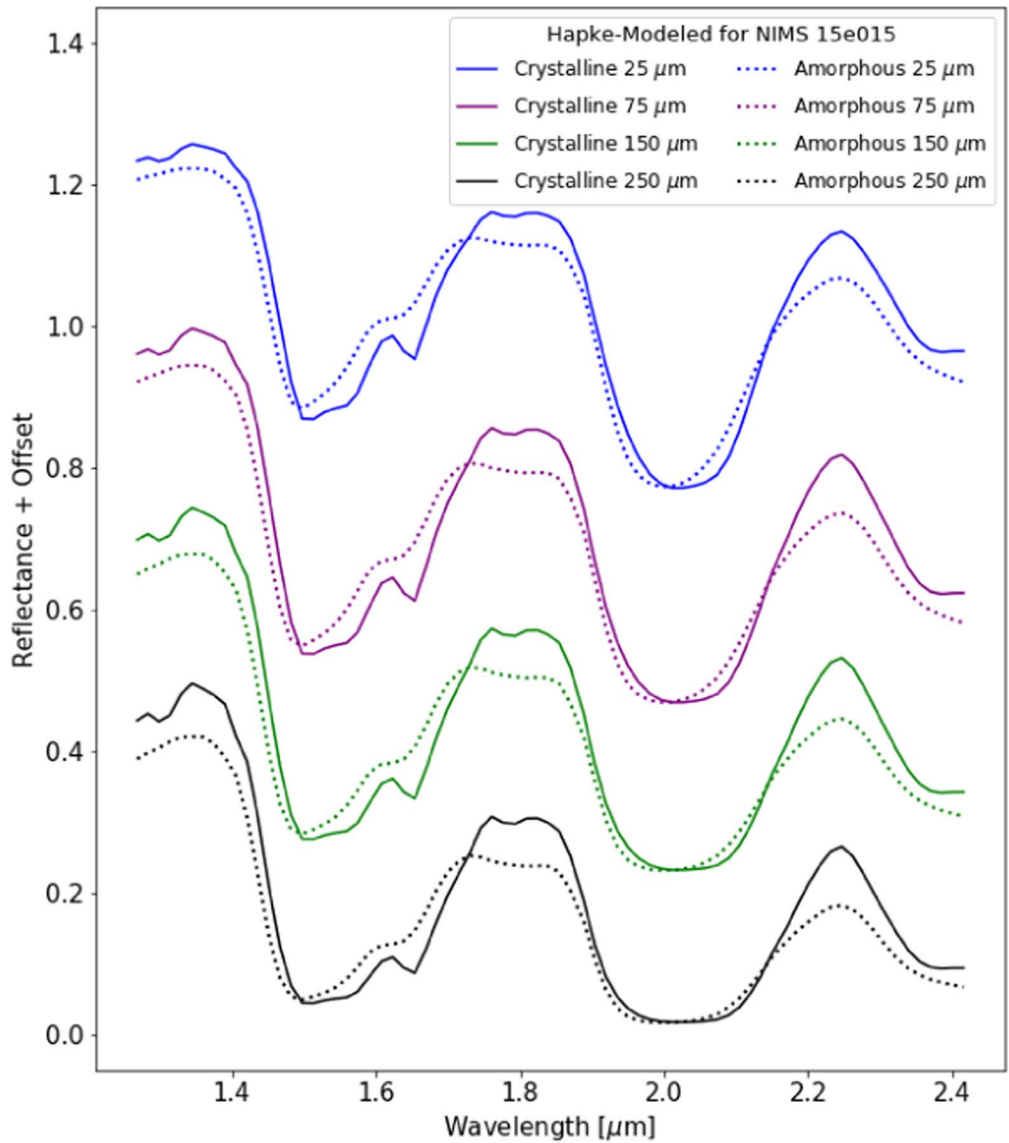

**Figure 2.** A subset of grain sizes of the Hapke-modeled amorphous and crystalline water-ice spectra produced for a viewing geometry that matches the location of the 15e015 NIMS cube and resampled at a spectral resolution that matches the 15e015 NIMS cube. Solid lines represent crystalline water-ice, whereas dashed lines represent amorphous water-ice. The spectra for each set of grain sizes are offset by 0.2 in reflectance for viewing clarity.

In order to determine the abundances of each of our endmembers in the NIMS image cubes, we performed a nonnegative least-squares (NNLS; Lawson & Hanson 1974; Chen & Plemmons 2009) spectral mixture analysis on each pixel separately. The NNLS algorithm solves the Karush–Kuhn–Tucker conditions (Karush 1939; Kuhn & Tucker 1951) for the least-squares problem and ensures that all constrained values are nonnegative, does not require that the total abundance of endmembers sums to 1.0 within a given pixel, and may choose a subset of the endmembers that results in the lowest $\chi^2$ without requiring the use of the full library of endmembers (Lawson & Hanson 1974). The benefit of not forcing the abundances to sum to 1.0 in each pixel is that it allows for the possibility of the presence of other materials that we have not included in this study that might exist on Europa's surface. Differences in viewing geometries and grain sizes of the reference material can also contribute to the uncertainty in the abundance total, so allowing a floating abundance total rather than a fixed abundance total ensures that the spectral mixture analysis process is not limited by the materials it is given.

We resampled all reference spectra to match the spectral resolution of the data set with the lowest spectral resolution, i.e., the NIMS spectra, at 0.0153 $\mu$m for the 15e015 cube and 0.0154 $\mu$m for the 17e009 cube. We tested the effects of different water-ice grain sizes and the presence of various water-ice spectra that are used in the endmember library to compute the lowest $\chi^2$ possible and therefore the best fit between the modeled and observed spectra.

### 3. Results and Discussion

#### 3.1. Derived Properties at the Equator

As a demonstration of the single-pixel spectral mixture analysis, Figure 4 displays examples of the model fits to individual pixel spectra, where pixels at ∼109° W, ∼113° W, and ∼116° W are displayed in order to sample three different locations across the 15e015 NIMS cube. As a comparison to the Dalton et al. (2012) results, we also provide their published model fits for the 15e015 NIMS cube, where the three panels do not necessarily correspond to the spectra provided in our three panels, but rather are shown as examples of their fitting process. Dalton et al. (2012) selected three types of spectra that were averaged together to produce the three spectra in Figure 4, Vis Bright (13 spectra), Vis Dark (24 spectra), and IR Bright (20 spectra). Our models in Figure 4 appeared by visual inspection to fit closer to the observed spectra than the fits from Dalton et al. (2012), despite the fact that our $\chi^2$ values are slightly larger. Since the overall fit did visually appear to be better (specifically between 1.7 and 1.8 $\mu$m, and at 1.95 $\mu$m), it is likely that our higher $\chi^2$ values were a result of the small noise spikes throughout our spectra, which are not present in the spectra from Dalton et al. (2012) owing to their multiple spectra averaging process.

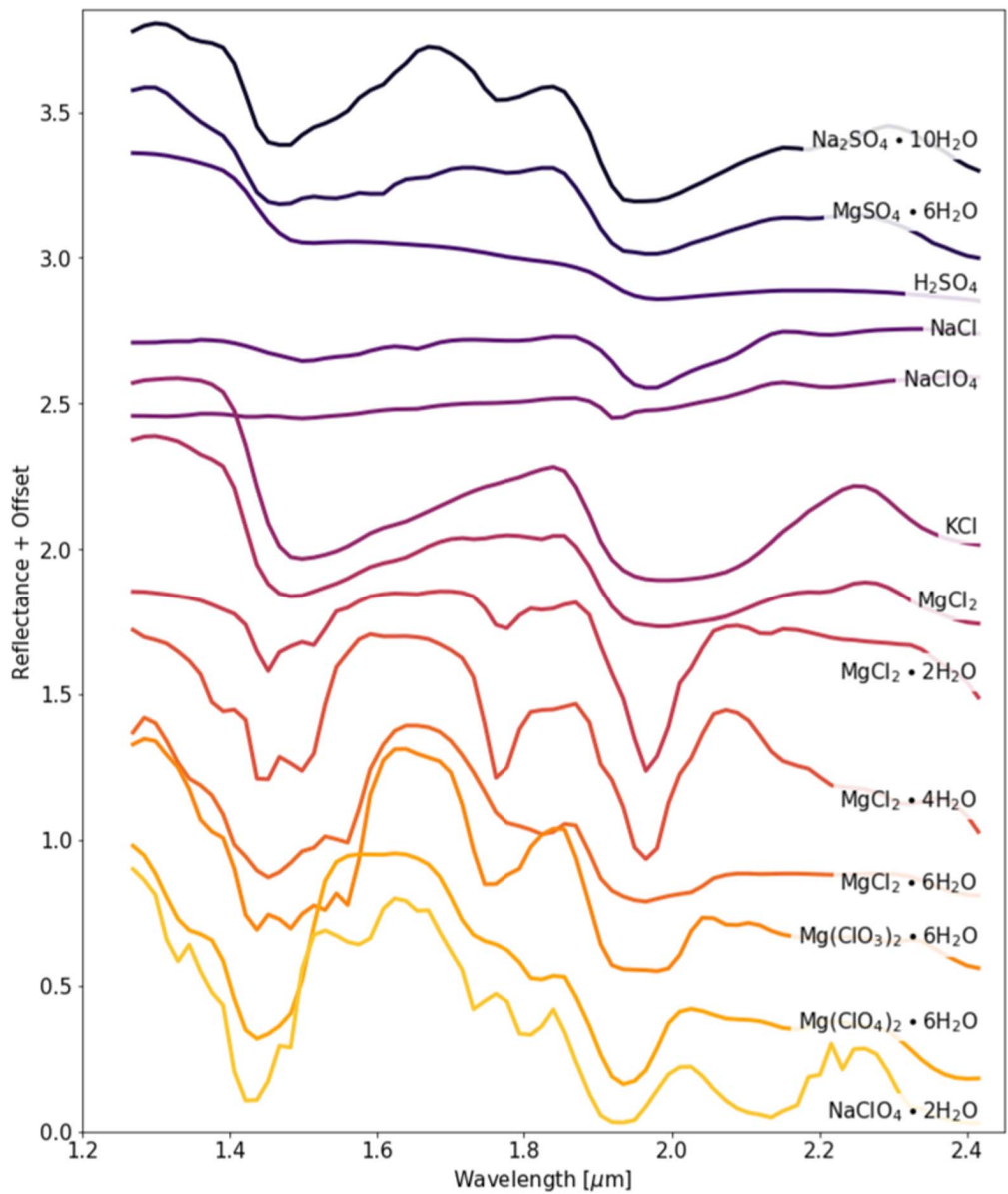

**Figure 3.** The full spectral reference library for non-water-ice materials resampled at a spectral resolution that matches that of the 15e015 NIMS image cube. These spectra were offset vertically for clarity.

Several lineae (including Drizzlecomb Linea) cross the 15e015 NIMS cube between 108° W and 112° W (Figure 5). Due to the presence of these lineae tracks, we predicted that the composition of the 15e015 NIMS cube would reflect alterations made to the surface due to subsurface material upwelling through surface lineae as a result of subsurface ice movement (Carlson et al. 2005). Lineae are associated with red, non-water-ice, hydrated material, though the question whether that material is of endogenic or exogenic origin is still an active discussion (Carlson et al. 2009). A long-standing debate exists regarding the source of surface materials on Europa; while magnesium and sodium brines may originate from the underlying ocean, sulfuric acid hydrate is likely radiolytically produced from interactions between deposited Io sulfur and Jupiter's magnetic field (Kargel 1991; Carlson et al. 2005). Determining the abundances and locations of these materials on the surface can contextualize the alteration processes occurring both below the surface and at the surface and aid in identifying whether that alteration is a result of endogenic or exogenic processing, or perhaps both. Furthermore, the distribution of crystalline and amorphous water-ice (and variations in grain sizes of each) across the 15e015 NIMS cube may provide further supporting information for surface alteration by exogenic or endogenic processing, as described in Section 1.

Table 2
Compound Names and Molecular Formulae for All Reference Materials Used in This Study

| Compound Name | Molecular Formula | Temperature (K) | Reference |
|---|---|---|---|
| Amorphous water-ice (synthetic) | $H_2O$ (type $I_a$) | 120 (15e015), | Mastrapa et al. (2008) |
| Crystalline water-ice (synthetic) | $H_2O$ (type $I_h$) | 90 (17e009) | |
| Magnesium chloride | $MgCl_2$ | 80 | Hanley et al. (2014) |
| | $MgCl_2 \cdot 2H_2O$ | | |
| | $MgCl_2 \cdot 4H_2O$ | | |
| | $MgCl_2 \cdot 6H_2O$ | | |
| Magnesium chlorate | $Mg(ClO_3)_2 \cdot 6H_2O$ | 80 | Hanley et al. (2014) |
| Magnesium perchlorate | $Mg(ClO_4)_2 \cdot 6H_2O$ | 80 | Hanley et al. (2014) |
| Hexahydrite (magnesium sulfate) | $MgSO_4 \cdot 6H_2O$ | 120 | Dalton (2003) |
| Mirabilite (sodium sulfate) | $Na_2SO_4 \cdot 10H_2O$ | 100 | Dalton et al. (2005) |
| Sodium chloride | $NaCl$ | 80 | Hanley et al. (2014) |
| Sodium perchlorate | $NaClO_4$ | 80 | Hanley et al. (2014) |
| | $NaClO_4 \cdot 2H_2O$ | | |
| Potassium chloride | $KCl$ | 80 | Hanley et al. (2014) |
| Sulfuric acid | $H_2SO_4 \cdot 8H_2O$ | 120 | Carlson et al. (2005) |

The high spatial resolution of the NIMS cubes (15e015: 3.0 km pixel$^{-1}$; 17e009: 1.5 km pixel$^{-1}$) and their structure being long, sporadic-data strips rather than maps (e.g., Figure 5) preclude displaying abundance maps here as physical longitude–latitude maps and would make distinguishing abundance representations from pixel to pixel quite difficult. Additionally, the image cubes do not extend substantially in the latitudinal direction (i.e., the 15e015 cube spans less than 0°.5 in latitude, and the 17e009 cube spans less than 1° in latitude), and analyzing the abundance variation with latitude did not yield any substantial results. For this reason, we assessed the results from the spectral mixture analysis in longitude space only; in other words, abundances for pixels at varying locations in latitude were plotted at the same longitude. It should be noted that the following longitudinal abundance profiles do not contain distinct latitudinal information, but all pixels in the original NIMS cube were accounted for and displayed within the figures. Shirley et al. (2010) conducted a similar style of "longitudinal gradient" test to understand how the abundances of species changed with longitude. While their study focused largely on NIMS observations at the leading–trailing hemisphere boundary, we performed a similar analysis to identify longitudinal gradients and correlations to surface features that may be present in these NIMS observations. Finally, we note that since the spatial resolution of the SSI imagery at both locations is much coarser than that of these NIMS observations, co-registering of the two data sets would be required in order to more explicitly identify which NIMS pixels might be affected by the lineae. We do not claim that any water/salt feature is a direct result of a specific linea, just that lineae exist in these cubes, and one or more could very well be contributing to the features, but higher-resolution imagery would be needed in order to assess a direct correlation between an abundance feature and a surface feature.

Results from the NNLS spectral mixture analysis are provided as longitudinal abundance profiles for crystalline water-ice (Figure 6), amorphous water-ice (Figure 7), and several of the dominant non-water-ice species (Figure 8) for the 15e015 NIMS cube. Large variations in water-ice and brine abundances at a given longitude are likely not real and represent the uncertainty associated with model selections based on the smallest $\chi^2$ alone, as well as the spectral resolution coarseness of the NIMS spectra. We therefore do not claim that the abundance profiles are exactly representative of the abundance distribution on Europa's surface, but they likely do represent general abundance trends in the NIMS cube. The minimum and maximum "goodness-of-fit" values, or $\chi^2$, of the 15e015 NIMS cube were 0.00753 and 0.09073, respectively. The mean and median $\chi^2$ of the 15e015 NIMS cube were 0.03633 and 0.03560, respectively. The distribution of $\chi^2$ across the 15e015 NIMS cube is shown in Figure 9. Dalton et al. (2012) calculated models with $\chi^2$ between 0.012653 and 0.021989 for their three averaged spectra across the 15e015 NIMS cube. While our $\chi^2$ values were slightly larger than theirs, we did retain spatial information and performed the spectral mixture analysis on each pixel within the cube, rather than averaged spectra of several types of regions.

Dalton et al. (2012) found that within the 15e015 NIMS cube, crystalline water-ice was exclusively present at 50 and 75 μm grain sizes. We found that a gradient exists across the cube; smaller grain sizes of crystalline water-ice (25–75 μm) are more abundant in the western longitudes (left side of Figure 6), whereas larger grain sizes are more prominent in the eastern longitudes, where several lineae cross through the NIMS cube (right side of Figure 6). This implies that crystalline water-ice grain sizes are smaller in regions where the surface has not undergone any recent alterations and has been exposed to radiation and sputtering over long timescales, which is not consistent with previous studies (Dalton et al. 2012) and is counterintuitive since sputtering causes larger grain sizes (Clark et al. 1983). However, it is possible that the longitudinal extent of the 15e015 NIMS cube is not great enough to demonstrate the effects of sputtering, and the grain size gradient is likely caused by an unidentified surface process.

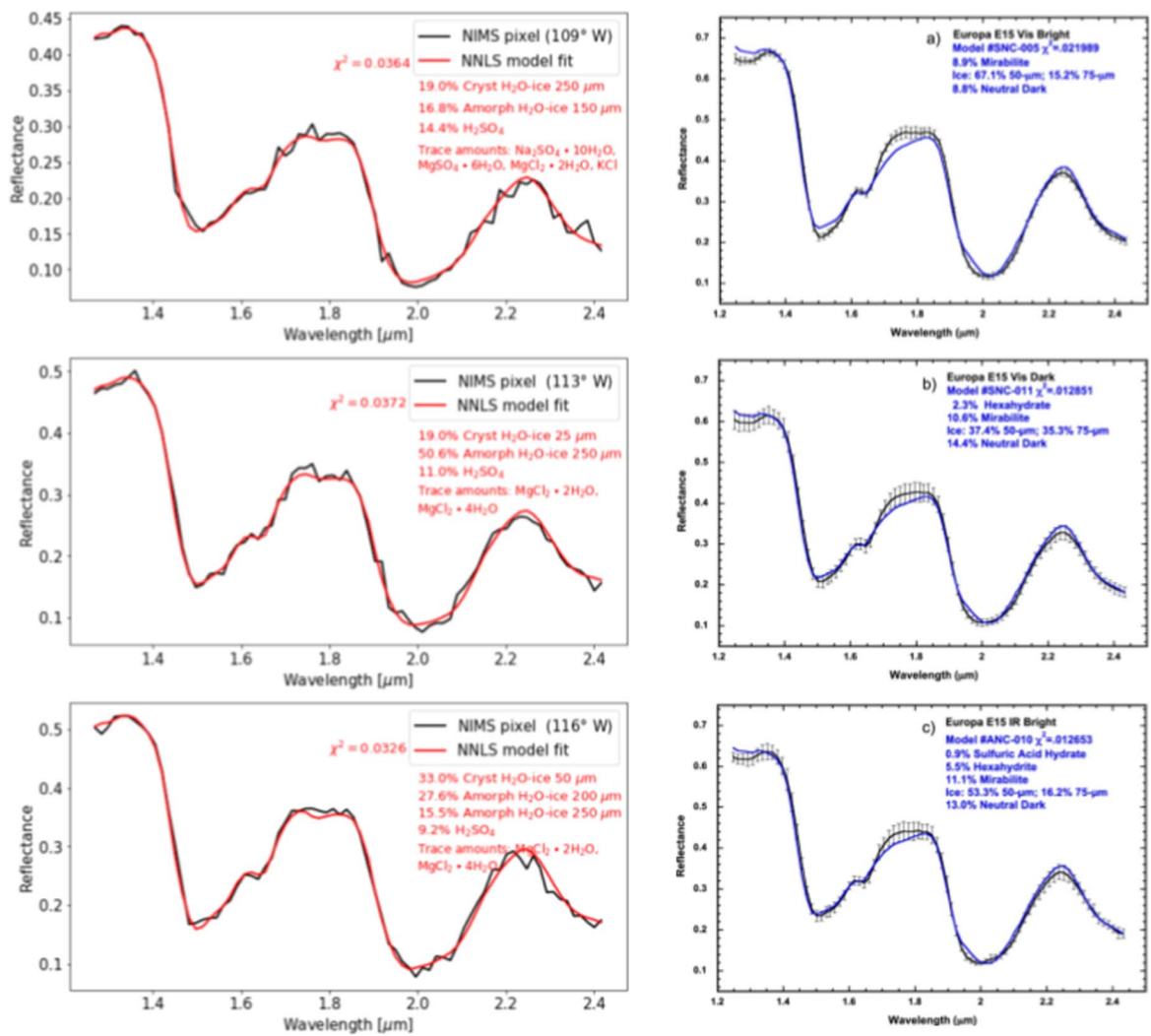

**Figure 4.** Left panels: NNLS spectral mixture analysis model results (red) for the spectra of three pixels (black) across the NIMS 15e015 cube. The $\chi^2$ of the model fit is provided for each panel at top center, and the approximate species abundances of the model fit are provided at center right. Species that are listed under "Trace amounts" are present in abundances of ∼5%. Right panels: linear spectral mixture analysis model results (blue) from three types of spectra (black) across the NIMS 15e015 cube (Figure 6 in Dalton et al. 2012). Dalton et al. (2012) selected three types of spectra that were averaged together to produce the above spectra, (a) Vis Bright (13 spectra), (b) Vis Dark (24 spectra), and (c) IR Bright (20 spectra).

While amorphous water-ice was not included in the Dalton et al. (2012) study, we found amorphous water-ice in a surprisingly high abundance in the 15e015 NIMS cube. We estimated a crystallinity of ∼35% based on the crystalline and amorphous water-ice abundances in the 15e015 NIMS cube (Figure 10). This is consistent with spectroscopically derived crystallinities of ∼30% for the full-disk leading hemisphere (Berdis et al. 2020).

Amorphous water-ice is not expected to be present over long timescales near the equator owing to the relatively short conversion timescale for water-ice to relax from the amorphous form to the crystalline form at local temperatures (∼100–130 K). Nevertheless, we found large grain sizes of amorphous water-ice (100–250 $\mu$m) at the equator. The presence of amorphous water-ice at the equator is likely only possible if local temperatures are much colder than previously thought, or if a plume eruption had recently deposited amorphous water-ice in this region shortly before observations were made. The latter event is not likely, because of the infrequency of plume eruption events at Europa (Sparks et al. 2016, 2017).

Of the magnesium-bearing species used in this study, we calculated that $MgSO_4 \cdot 6H_2O$ was the most abundant in the 15e015 NIMS cube, with trace amounts of $MgCl_2 \cdot 2H_2O$ and $MgCl_2 \cdot 4H_2O$. This is consistent with Dalton et al. (2012), who identified $MgSO_4 \cdot 6H_2O$ in the 15e015 NIMS cube in roughly similar quantities, i.e., less than ∼10%. Near 113.5° W, we found a smooth decrease from ∼9% abundance at 113.2° W to 0% abundance at 113.5° W and a smooth increase back to ∼9% abundance at 113.8° W, which is statistically significant. The cause of this feature is unknown, as there does not appear to be a geological feature at this location (Figure 5). Buratti & Golombek (1988) and Carlson et al. (1999) referred to the presence of features that "show bright central bands flanked by bands of darker material" as "triple bands," a surface feature that could potentially produce this pattern in the longitudinal abundance profile for $MgSO_4 \cdot 6H_2O$ (Figure 8). Higher-resolution imagery (such as from the Europa Clipper mission currently planned to launch in 2024, or a future lander mission) is needed to identify the source for this double- or triple-banded surface feature.

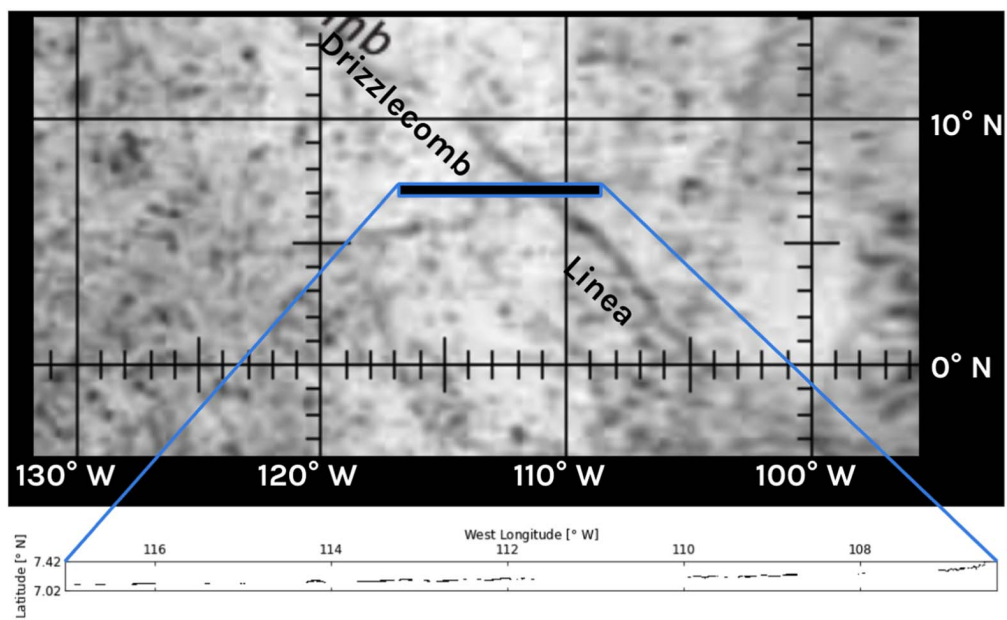

**Figure 5.** Representation of the 15e015 NIMS cube overlaying the relevant region from an image mosaic (image courtesy of USGS Astrogeology Science Center; Controlled Photomosaic Map of Europa, Je 15M CMN). The black rectangle with a blue border represents the approximate location and size of the 15e015 NIMS cube. Drizzlecomb Linea is labeled and crosses through the NIMS cube; however, the exact location where it crosses the cube is unknown, as the two data sets are not co-registered. Below the image is a spatial latitude–longitude representation of the data that exists within the 15e015 cube. The black pixels represent locations where image spectroscopy data were acquired.

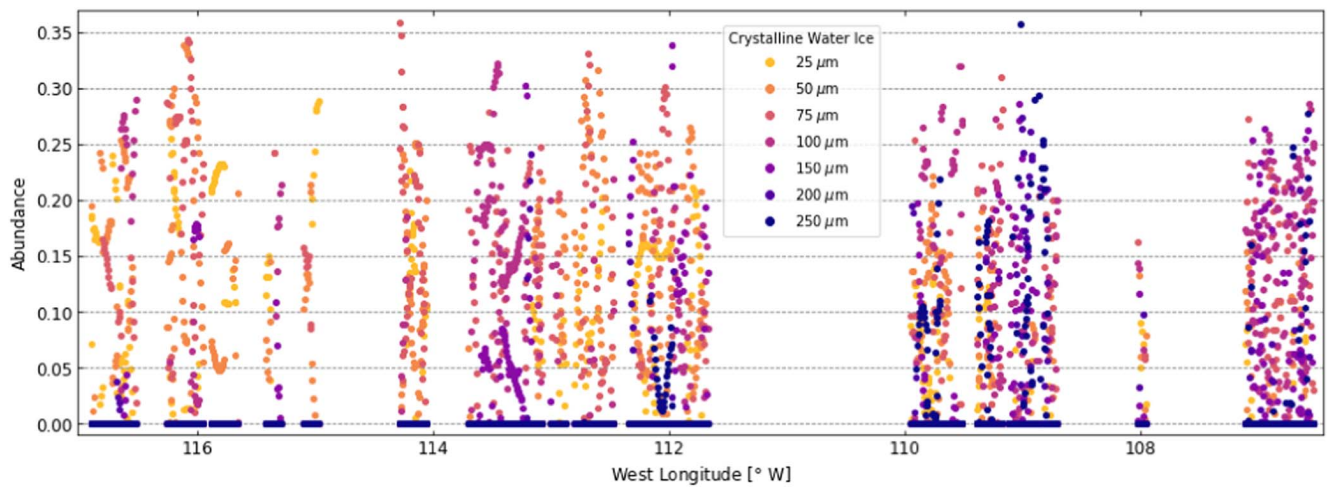

**Figure 6.** Abundance profiles across the 15e015 NIMS cube for crystalline water-ice at grain sizes ranging from 25 to 250 $\mu$m.

This surface feature located at 113.2°–113.8° W was also present in the abundance profile of the sodium-bearing species. An enhancement existed in $Na_2SO_4 \cdot 10H_2O$ between 113.2° W and 113.5° W, corresponding to one of the $MgSO_4 \cdot 6H_2O$ peaks in the hypothesized double-banded surface feature. An increase in $Na_2SO_4 \cdot 10H_2O$ in the center of an $MgSO_4 \cdot 6H_2O$ valley may indicate subsurface processes that are transporting sodium-rich material to the surface through a crack feature (Hand & Carlson 2015; Trumbo et al. 2019); however, it is unclear what an enhancement on the edge of a double-banded surface feature may indicate. Furthermore, the subtle enhancement of $Na_2SO_4 \cdot 10H_2O$ and KCl at 109° W is likely caused by upwelled material through lineae, of which there are several in the eastern longitudes of the 15e015 NIMS cube.

### 3.2. Derived Properties near the South Pole

Similar to the single-pixel spectral fits in the previous section, Figure 11 displays examples of the model fits to individual pixel spectra, where pixels at ∼118° W, ∼124° W, and ∼127° W are displayed in order to sample three different locations across the 17e009 NIMS cube. We also provide published model fits from Dalton et al. (2012) for the 17e009 NIMS cube. Similar to the spectral fit comparison for the 15e015 NIMS cube, it is likely that our higher $\chi^2$ values were a result of the small noise spikes throughout our spectra, which are not present in the spectra from Dalton et al. (2012) owing to their multiple spectra averaging process.

Although the spatial resolution of the SSI imagery at the location of the 17e009 NIMS cube was generally quite low (Figure 12), at least one geological feature could be identified; Adonis Linea crosses the NIMS observation at ∼124°–125° W.

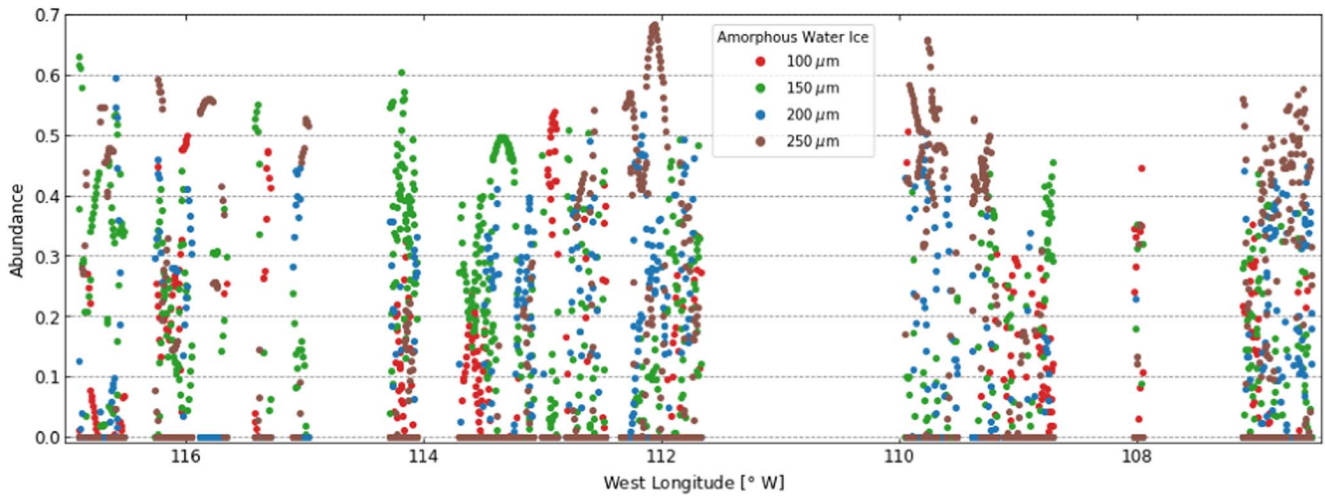

**Figure 7.** Abundance profiles across the 15e015 NIMS cube for amorphous water-ice at grain sizes ranging from 100 to 250 $\mu$m, where grain sizes less than 100 $\mu$m are not included, as they were present in trace amounts.

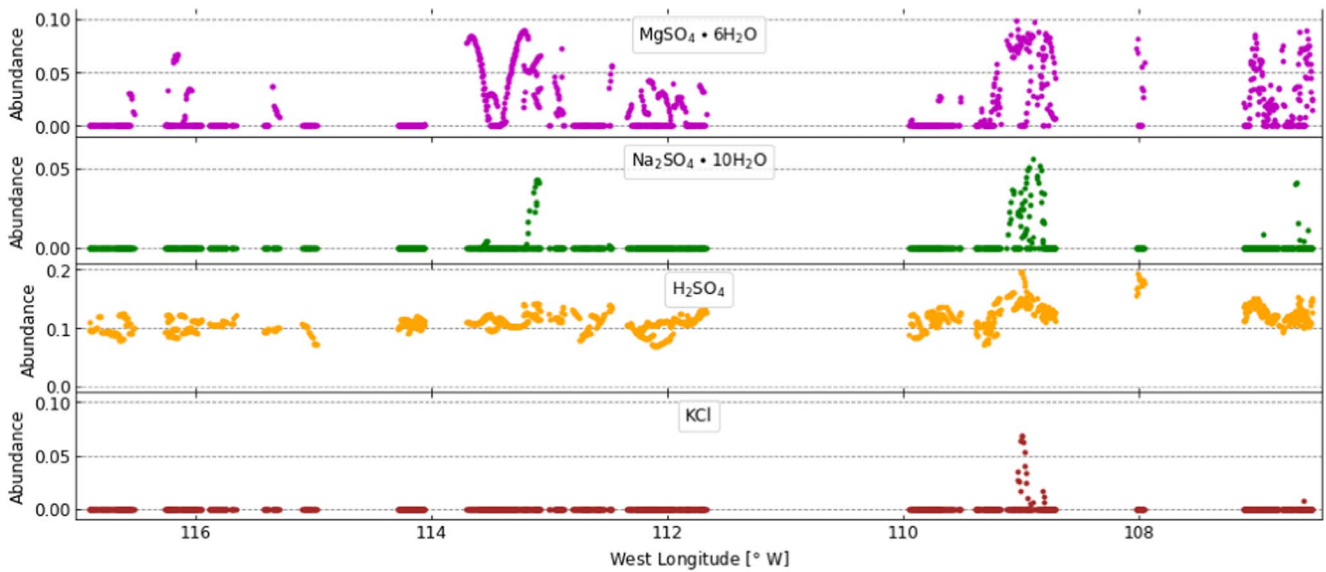

**Figure 8.** Abundance profiles across the 15e015 NIMS cube for $MgSO_4 \cdot 6H_2O$, $Na_2SO_4 \cdot 10H_2O$, $H_2SO_4$, and KCl; all other non-water-ice species are not included, as they were present in trace amounts.

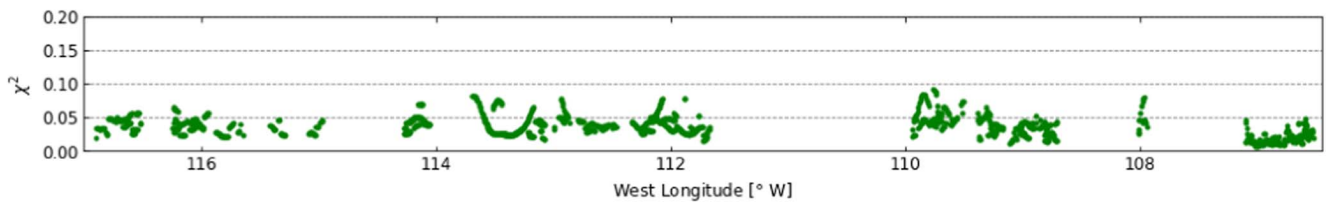

**Figure 9.** Distribution of $\chi^2$ across the NIMS 15e015 NIMS cube. The mean $\chi^2$ of pixels in which the NNLS was performed is 0.03633.

As discussed in Section 3.1, we predicted that the composition of the 17e009 NIMS cube would reflect alterations made to the surface by the possible upwelling of subsurface material through cracks and surface lineae.

Similar to the 15e015 NIMS cube, the 17e009 NIMS cube had a high spatial resolution and more resembled a longitudinal strip rather than a map, where only a single ground track of data was acquired (Figure 12). As discussed in Section 3.1, displaying abundance maps as longitude–latitude maps made distinguishing color map colors from pixel to pixel quite difficult, where each pixel was visually quite small. The 17e009 NIMS cube spans slightly more than 0°.5 in latitude, so abundance variations with latitude were not substantially revealing. We display 17e009 NIMS cube abundance profiles across longitude in the same way that the 15e015 NIMS cube abundance profiles are displayed in Figures 6–8. Longitudinal abundance profiles are provided for crystalline water-ice (Figure 13), amorphous water-ice (Figure 14), and several of the dominant non-water-ice species (Figure 15) for the 17e009 NIMS cube.

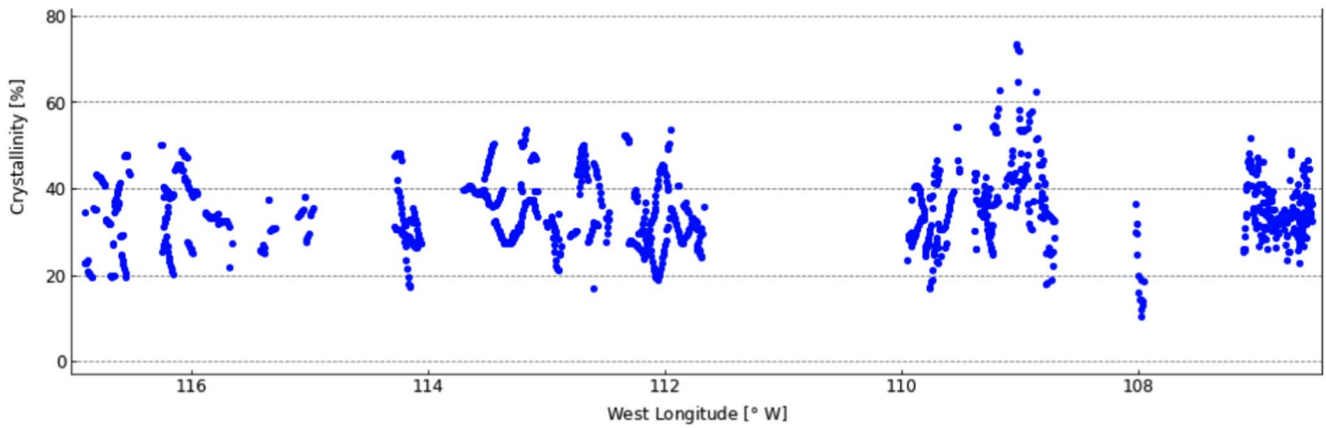

**Figure 10.** Distribution of crystallinity across the 15e015 NIMS cube. The mean crystallinity of all pixels in this cube is ∼35%.

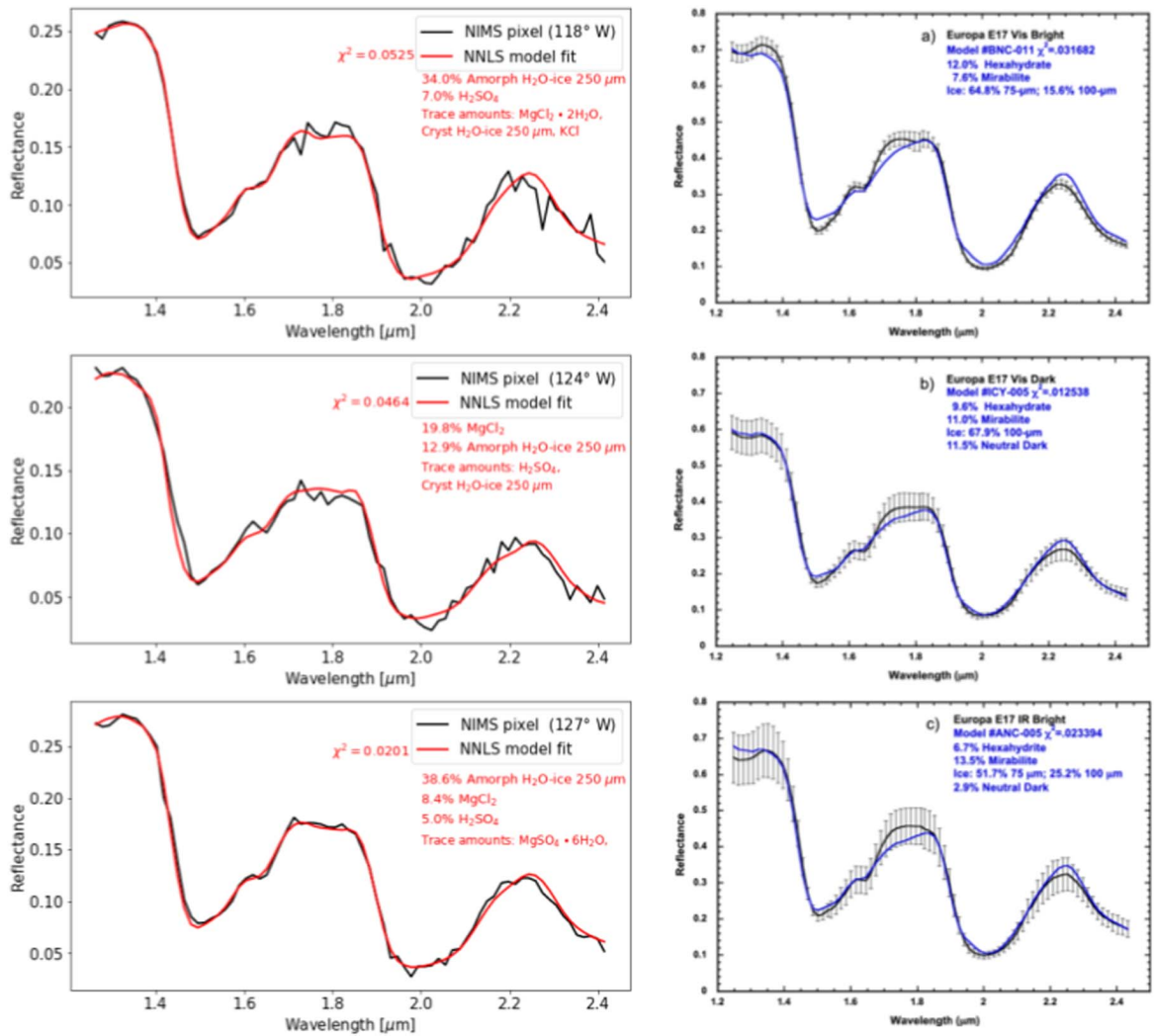

**Figure 11.** Left panels: NNLS spectral mixture analysis model results (red) from the spectra of three pixels (black) across the NIMS 17e009 cube. The $\chi^2$ of the model fit is provided for each panel at top center, and the approximate species abundances of the model fit are provided at center right. Species that are listed under "Trace amounts" were present in abundances of ∼5%. Right panels: linear spectral mixture analysis model results (blue) from three types of spectra (black) across the NIMS 17e009 cube (Figure 8 in Dalton et al. 2012). Dalton et al. (2012) selected three types of spectra that were averaged together to produce the above spectra, (a) Vis Bright (13 spectra), (b) Vis Dark (24 spectra), and (c) IR Bright (20 spectra).

The minimum and maximum "goodness-of-fit" values, or $\chi^2$, of the 17e009 NIMS cube were 0.00730 and 0.12045, respectively. The mean and median $\chi^2$ of the 17e009 NIMS cube were 0.03285 and 0.02878, respectively. Dalton et al. (2012) calculated models with $\chi^2$ between 0.012538 and 0.031682 for their three averaged spectra across the 17e009 NIMS cube. Similar to the 15e015 NIMS cube, our $\chi^2$ values were slightly larger than theirs; however, we did retain spatial information and performed the spectral mixture analysis on each pixel within the cube, rather than averaged spectra of several types of regions. The distribution of $\chi^2$ across the 17e009 NIMS cube is shown in Figure 16.

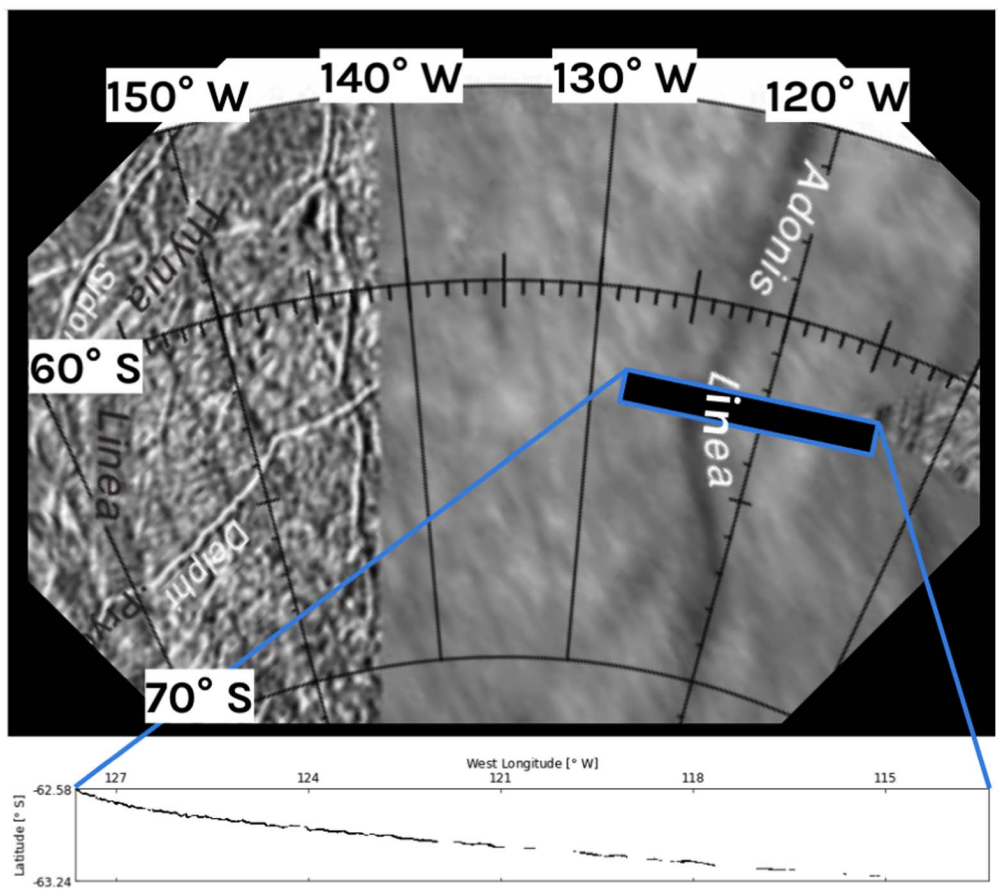

**Figure 12.** Representation of the 17e009 NIMS cube overlaying the relevant region from an image mosaic (image courtesy of USGS Astrogeology Science Center; Controlled Photomosaic Map of Europa, Je 15M CMN). The black rectangle with a blue border represents the approximate location and size of the 17e009 NIMS cube. Adonis Linea is labeled and crosses through the 17e009 NIMS cube; however, the exact location where it crosses the cube is unknown, as the two data sets are not co-registered. Below the image is a spatial latitude–longitude representation of the data that exists within the 15e015 cube. The black pixels represent locations where image spectroscopy data were acquired.

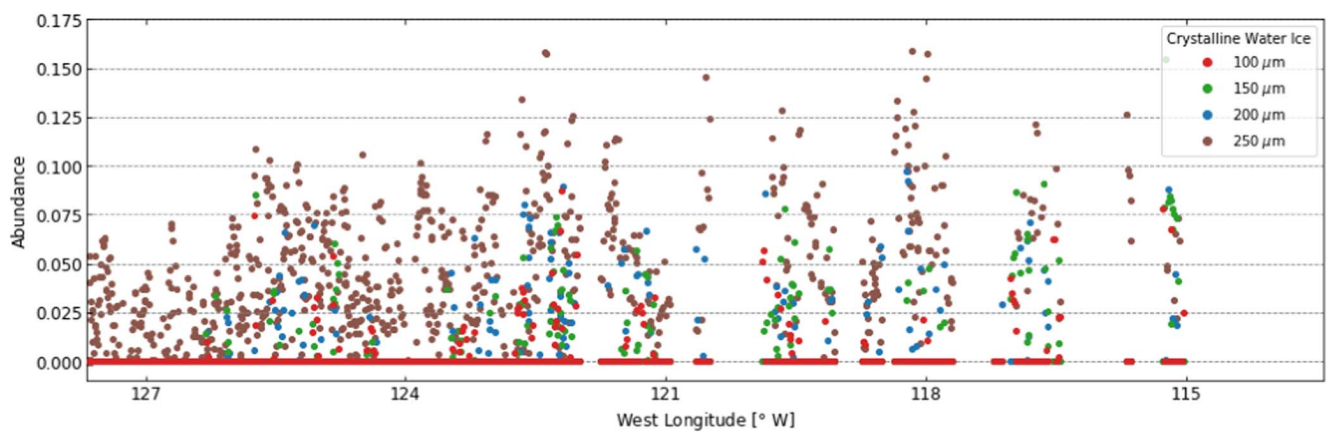

**Figure 13.** Abundance profiles across the 17e009 NIMS cube for crystalline water-ice at grain sizes ranging from 100 to 250 $\mu$m, where grain sizes less than 100 $\mu$m are not included, as they were present in trace amounts.

While Dalton et al. (2012) found larger grain sizes of crystalline water-ice in the 17e009 NIMS cube compared to the 15e015 NIMS cube, they found mostly 75 and 100 $\mu$m grain sizes, and no detectable amount of 250 $\mu$m grain sizes. We find that grain sizes of 200 and 250 $\mu$m for both crystalline and amorphous water-ice were most abundant in the 17e009 NIMS cube. Amorphous water-ice appears to dominate this region, with abundances of ~40%, compared to crystalline water-ice with abundances of ≲10%. For the 15e015 NIMS cube, we identified a trend in the crystalline water-ice grain size distribution across the cube as longitude decreased; smaller grain sizes of crystalline water-ice were more abundant in the western longitudes, whereas larger grain sizes were more prominent in the eastern longitudes. However, no such grain size gradient existed for either crystalline or amorphous water-ice within the 17e009 NIMS cube.

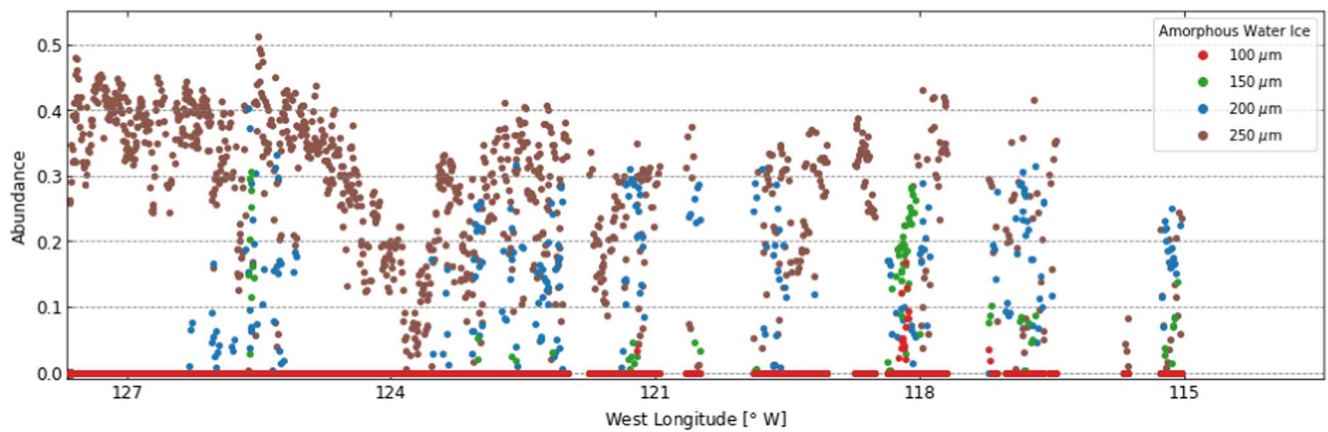

**Figure 14.** Abundance profiles across the 17e009 NIMS cube for amorphous water-ice at grain sizes ranging from 100 to 250 $\mu$m, where grain sizes less than 100 $\mu$m are not included, as they were present in trace amounts.

We estimated a crystallinity of ~15% based on the crystalline and amorphous water-ice abundances in the 17e009 NIMS cube (Figure 17). This was slightly lower than, but still consistent with, spectroscopically derived crystallinities of ~30% for the full-disk leading hemisphere, which is dominated by warmer (and therefore more crystalline), lower-latitude locations (Berdis et al. 2020). As discussed previously, Adonis Linea crosses the NIMS observation at ~124°–125° W. We identified a feature in the 250 $\mu$m grain size of amorphous water-ice; a gradual but noticeable dip in abundance occurred between ~124° and 125° W, with a minimum abundance occurring around 124° W. No change (either positive or negative) of the abundance of crystalline water-ice was detected at this location.

The 124°–125° W location where Adonis Linea crosses the 17e009 NIMS cube is correlated with changes in the $H_2SO_4$, $MgCl_2$ abundances, and 250 $\mu$m amorphous water-ice (Figure 15). At 124° W, the $MgCl_2$ abundance formed a double-peak feature; a single-peak feature also existed at 121.5° W. These two features are mirrored in the $H_2SO_4$ abundance profile, where increases in the $MgCl_2$ abundances correlate to decreases in the $H_2SO_4$ abundances; a double-trough feature occurred at 124° W and a single-trough feature occurred at 121.5° W in the $H_2SO_4$ abundance profile. We hypothesize that these locations correlate to the locations of bands that have recently undergone alteration in the form of $MgCl_2$-abundant material upwelling from below the surface and, in the process, reducing the signal of radiolytically produced $H_2SO_4$ that was present on the surface. This is consistent with previous results; Carlson et al. (2005) found that the abundance of $H_2SO_4$ was lower within lineae compared to the nearby material. The double-peak structure could arise from a double ridge or triple band surface feature (Figure 15), though it is difficult to identify the structure of Adonis Linea at this location owing to the low resolution of the SSI imagery. Higher-resolution imaging of the surface at this location is needed to assess the correlation of double ridge structures with the abundances of $H_2SO_4$ and $MgCl_2$.

### 3.3. Location Comparison of Composition and Crystallinity

We performed NNLS spectral mixture analyses on high spatial resolution NIMS cubes (or longitudinal "strips") in order to identify how surface water-ice crystallinity is influenced by physical processing. We included hydrated sulfuric acid and brines in our investigation not only to ensure a more complete spectral reference library but also to also search for additional evidence of physical processing. Here we discuss the results from the non-water-ice materials (magnesium-bearing species, sodium-bearing species, $H_2SO_4$, KCl, and the dark absorber, in that order), and then we discuss the water-ice crystallinity results.

$MgSO_4 \cdot 6H_2O$ exists at both locations in similar abundances, roughly 5% (Figure 8). Whereas a clear feature exists in the longitudinal abundance of $MgSO_4 \cdot 6H_2O$ at the equator that may be linked to the presence of a crossing linea, no such longitudinal abundance feature exists in the south polar observations. However, $MgCl_2$ does trace a hypothesized double ridge or triple band surface feature at approximately the location where Adonis Linea crosses the NIMS observation. The fact that two different magnesium-bearing species are related to surface features at two different locations on the surface indicates either that there are two different processes that caused these features or that the subsurface material is inherently different at the equator compared to the south pole. Higher spatial resolution imagery from a future instrument, such as Europa Clipper's Europa Imaging System (EIS; Patterson et al. 2019), is needed in order to determine whether the surface features that cross these two NIMS observations are formed by inherently different processes. The detection of Adonis Linea based solely on brine composition abundance profiles demonstrates that the analysis described in this study could provide a method for determining which surface features are present in regions where the resolution of the imagery is too low to discern geological features and can inform region-of-interest observation selections for the Europa Clipper and Europa Lander missions.

Sodium-bearing species are present in negligible abundances ($\lesssim 5\%$) within both NIMS observations; however, a subtle enhancement in $Na_2SO_4 \cdot 10H_2O$ near 109° W in the equatorial 15e015 NIMS cube (Figure 8) may relate to lineae or other surface features. This lack of sodium-bearing species is surprising, as Dalton et al. (2012) found sodium-bearing species (specifically

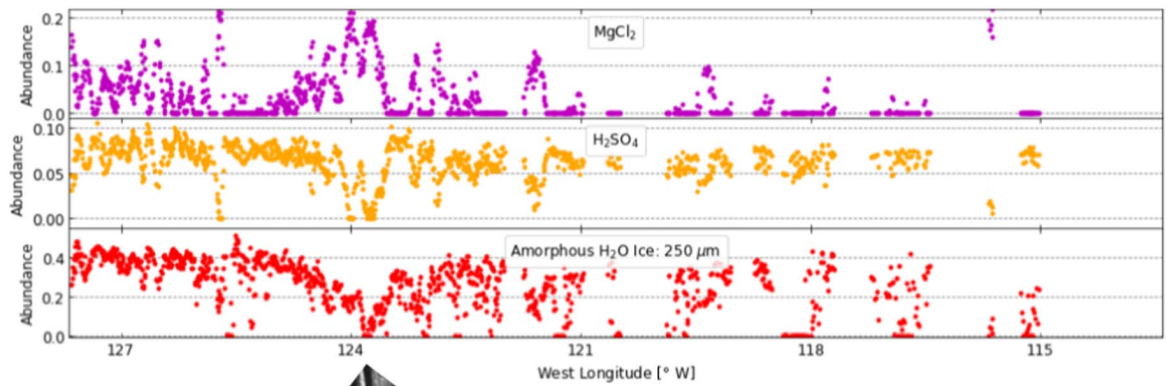

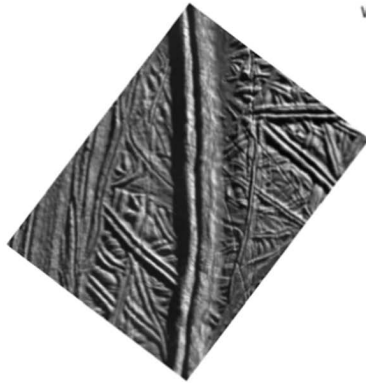

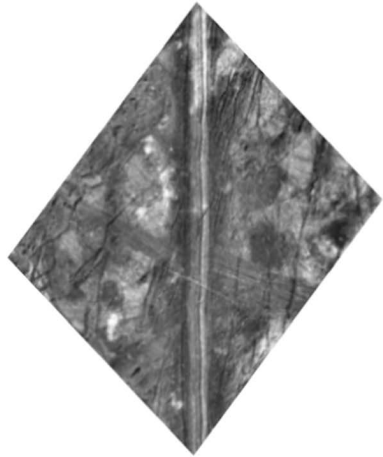

**Figure 15.** Abundance profiles across the 17e009 NIMS cube for $MgCl_2$, $H_2SO_4$, and 250 $\mu$m amorphous water-ice; all other non-water-ice species are not included, as they were present in trace amounts. These three species display a feature located at ∼124°–125° W and are likely influenced by a geological feature at that location. Images of a doublet ridge and triple band are shown to demonstrate possible surface features that could cause these abundance patterns. The doublet ridge pictured is ∼2 km wide, and the triple band pictured is ∼10 km wide; however, widths for both types of features vary substantially across Europa's surface. Images courtesy of Fagents et al. (2000).

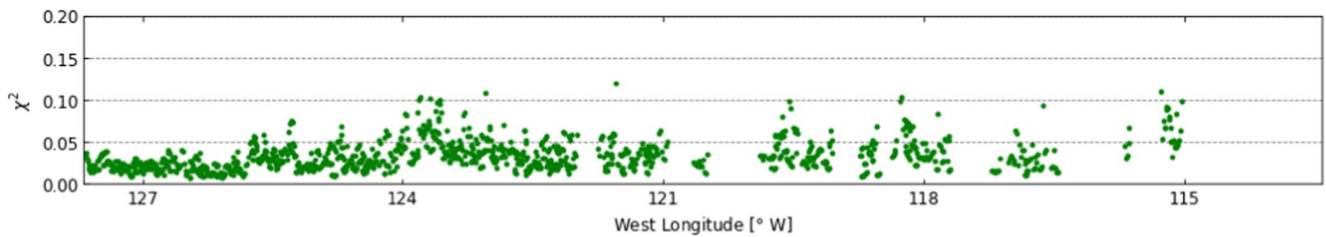

**Figure 16.** Distribution of $\chi^2$ across the NIMS 17e009 NIMS cube. The $\chi^2$ appears to improve at higher longitudes; the mean $\chi^2$ of pixels in which the NNLS was performed is 0.03285.

$Na_2SO_4 \cdot 10H_2O$) in nonnegligible quantities within both the equatorial and south polar NIMS cubes. Furthermore, Trumbo et al. (2019) detected NaCl using the Hubble Space Telescope (HST) Space Telescope Imaging Spectrograph (STIS) in nonnegligible quantities, especially near the equator of the leading hemisphere, where geologically disrupted chaos terrain is widespread, suggesting that sodium-bearing species have an endogenic origin (Trumbo et al. 2019). The minimal detection of sodium-bearing species in our study suggests that these species exist in regions that have had more significant surface disruption, such as chaos terrain, and do not exist in large quantities near lineae.

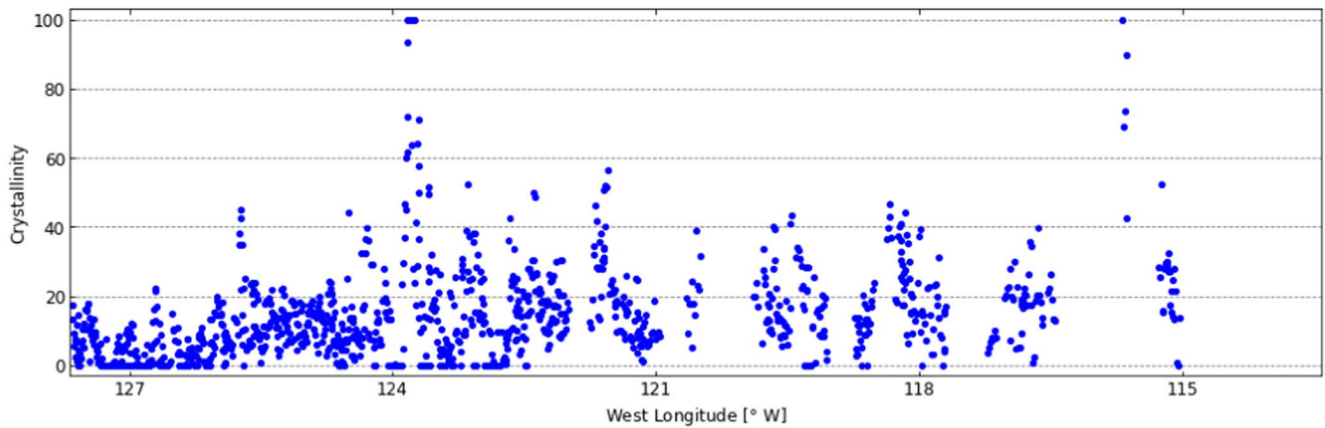

**Figure 17.** Distribution of crystallinity across the 17e009 NIMS cube. The mean crystallinity of all pixels in this cube is ∼15%.

We detected sulfuric acid hydrate ($H_2SO_4 \cdot 8H_2O$) in abundances of ∼5%–15% within our NIMS observations (Figures 8 and 15), which is consistent with previous estimates of ∼10% on the leading hemisphere (Carlson et al. 2005). We found an overall higher abundance of sulfuric acid hydrate near the equator compared to the south pole. This is also consistent with previous results; a "bulls-eye" distribution of sulfuric acid hydrate was identified at the apex of the trailing hemisphere and may influence a similar pattern on the leading hemisphere (Carlson et al. 2005).

While potassium chloride (KCl) exhibits very similar spectral features to water-ice in the near-IR (NIR), the NNLS algorithm detected negligible abundances of KCl in both NIMS observations. This is inconsistent with previous results, which predict that KCl should exist in large quantities on the leading hemisphere (Brown & Hand 2013). However, Brown & Hand (2013) noted that these species display distorted water bands with no distinct spectral features when hydrated, making them difficult to detect spectroscopically, and perhaps cannot be distinguished from water-ice at the NIMS spectral resolution.

The dark absorber endmember was present only rarely in the model fits to the NIMS cubes and exhibited a nonzero abundance for ∼5 pixels in each NIMS observation. This is unusual, as Dalton et al. (2012) found nonnegligible amounts of the dark absorber in their spectral mixture analysis. Our $\chi^2$ values were slightly larger than theirs, and perhaps a forced inclusion of a larger fraction of the dark absorber would produce a model fit with a similar $\chi^2$ value. However, our NNLS algorithm selected larger grain sizes of amorphous and crystalline water-ice compared to the fits modeled in Dalton et al. (2012); since the presence of larger grain sizes of water-ice can decrease the overall reflectance of the spectra, it is likely that the selection of larger grain sizes by the NNLS algorithm diminished the need to select the dark absorber.

As described previously, crystalline water-ice is expected to be more abundant near the equator owing to warmer temperatures, and amorphous water-ice is expected to be more abundant near the poles owing to sublimation of ice and/or migration of plume material (Kouchi et al. 1994;

Hansen & McCord 2004). We found that comparatively there was more crystalline water-ice at the equator than at the south pole (Figures 6 and 13). However, while fractionally there were fewer pixels selected as containing amorphous-water-ice ≳5% abundance at the equator compared to the pole, there was a higher abundance of amorphous water-ice within those selected pixels at the equator. The higher overall abundance of amorphous water-ice over crystalline water-ice in both cubes could be a result of the depth into the surface that is probed by NIR observations. Hansen & McCord (2004) suggested that deeper than ∼1 mm (i.e., the depth probed by NIR observations) water-ice on Europa is predominantly crystalline; however, nearly all of the water-ice above a depth of ∼1 mm is amorphous. While warm temperatures near the equator could be thermally relaxing the thin topmost monolayer into crystalline water-ice, the subsurface layers (up to a depth of ∼1 mm) may not be receiving enough solar insolation to relax into the crystalline form, allowing the water-ice to be predominantly amorphous.

However, the described scenario does not explain why there would be more amorphous water-ice at the equator than at the south pole. One cause for this distribution of amorphous water-ice across the surface is location-dependent radiation processing. While the trailing hemisphere receives a majority of the charged particle bombardment at Europa, the leading hemisphere likely also receives a significant (though smaller) dose of high-energy flux of electrons, protons, and ions (Paranicas et al. 2001; Patterson et al. 2012; Nordheim et al. 2018). The particle flux on both the leading and trailing hemispheres exhibits bulls-eye or target patterns centered on each hemisphere's apex; therefore, the poles receive a lower dose of charged particle bombardment than the equator (Nordheim et al. 2018). Results from our study indicate that there is more radiation-altered amorphous water-ice existing at the equator up to a depth of ∼1 mm than previously thought, and that the abundance of amorphous water-ice surpasses that of crystalline water-ice at the equator, suggesting that radiation alteration is overpowering the efficiency of thermal relaxation converting amorphous into crystalline water-ice.

Alternatively, the determination of a higher abundance of amorphous water-ice at the equator and south pole may also be a result of the coarse spectral resolution of the NIMS instrument. Although the NNLS algorithm is capable of correctly discerning between crystalline and amorphous water-ice at the spectral resolution of the NIMS instrument, the NNLS algorithm would return lower $\chi^2$ values, and in theory provide better model fits, if it were given data with a higher spectral resolution. The spectral resolution of the NIMS Long Spectrometer Mode data is ∼0.015 μm; Europa Clipper's Mapping Imaging Spectrometer for Europa (MISE) will have a spectral resolution of 0.010 μm and could acquire data at a spectral resolution of 0.005 μm if a Nyquist sampling rate is offered for a similar Long Spectrometer Mode. Finer spectral resolution data from this future mapping spectrometer is needed in order to identify whether this overabundance of amorphous water-ice at the equator compared to the south pole is real or a result of coarse spectral resolution.

To summarize the crystallinity results presented in this study, the spatial distribution of crystallinity across the two NIMS cubes (Figures 10 and 17) does not change a substantial amount as a function of longitude; however, there is a crystallinity enhancement in the south pole 17e009 NIMS cube located roughly where Adonis Linea crosses the cube near 124° W. As discussed previously, this feature is correlated with a reduced abundance of amorphous water-ice, with a dependency between the $H_2SO_4$ and $MgCl_2$ abundances in the form of a hypothesized double ridge or triple band feature on the surface. The increased crystallinity at this location does not appear to be due to an increase in crystalline water-ice (Figure 13) but rather only the decrease in amorphous water-ice, so there is no evidence at this location for partial warming due to upwelled warmer subsurface material.

The mean crystallinity of the 15e015 NIMS cube was ∼35%, and the mean crystallinity of the 17e009 NIMS cube was ∼15%. A lower crystallinity near the pole is expected owing to (1) colder temperatures, which indicates a longer amount of time for the amorphous water-ice to crystallize (Kouchi et al. 1994; Jenniskens et al. 1998; Mastrapa et al. 2013), and (2) migration of amorphous plume material (Kouchi et al. 1994; Hansen & McCord 2004). These crystallinities are in agreement with those derived using full-disk ground-based spectroscopic observations of Europa's leading hemisphere and laboratory-produced crystalline and amorphous water-ice (Berdis et al. 2020). They found that the spectroscopically derived crystallinity of Europa's full-disk leading hemisphere is ∼27%–36%, whereas the crystallinity derived from thermo-physical modeling and particle flux is ∼80%–95%. The calculated crystallinity as derived spectroscopically is therefore independent of the methodology used, i.e., Berdis et al. (2020) made use of the 1.65/1.5 μm band area ratios, whereas this study implemented a spectral mixture analysis on the full 1.2–2.4 μm range of the NIR spectrum. This therefore provides a check on the spectroscopically derived crystallinities calculated in Berdis et al. (2020) and strengthens the argument that particle flux and thermophysical modeling alone cannot reproduce the crystallinity present on Europa's leading hemisphere, supporting their hypothesis that there are additional processes occurring on Europa's surface that were not included in their model.

## 4. Conclusions

We implemented a spectral mixture analysis on two Galileo NIMS hyperspectral image cubes located near the equator (cube 15e015) and south pole (cube 17e009), both on the leading hemisphere, to derive the crystallinity and brine composition of Europa's surface. We estimated a mean crystallinity of ∼35% in the equatorial 15e015 NIMS cube and a mean crystallinity of ∼15% in the south polar 17e009 NIMS cube. This is in agreement with previous literature that suggests that a lower crystallinity near the pole is expected owing to (1) colder temperatures, which indicate a longer amount of time for the water-ice to crystallize (Kouchi et al. 1994; Jenniskens et al. 1998; Mastrapa et al. 2013), and (2) migration of amorphous plume material (Kouchi et al. 1994; Hansen & McCord 2004). These crystallinity values are also in agreement with those derived using full-disk ground-based spectroscopic observations of Europa's leading hemisphere and laboratory-produced crystalline and amorphous water-ice (∼30%; Berdis et al. 2020).

We found predominantly large grain sizes of amorphous water-ice (100–250 μm) at the equator; however, amorphous water-ice is not expected to be present over long timescales near the equator as a result of the relatively short conversion timescale for water-ice to relax from the amorphous form to the crystalline form at temperatures that are expected near the equator (∼100–130 K). A higher abundance of amorphous water-ice compared to crystalline water-ice at the equator suggests that radiation alteration is overpowering the efficiency of thermal relaxation converting amorphous into crystalline water-ice, which is in agreement with Hansen & McCord (2004) but conflicts with the results from Ligier et al. (2016).

We identified a decrease in abundance of the 250 μm grain-sized amorphous water-ice that corresponded with the location at which Adonis Linea crossed the south polar NIMS observation. We hypothesized that regions with enhanced or reduced water-ice or brine abundances correlate to the locations of lineae that have recently undergone alteration. This feature in the amorphous water-ice abundance also corresponded to a double-peak structure in the $MgCl_2$ and $H_2SO_4$ abundances, where increases in the $MgCl_2$ abundances correlated with decreases in the $H_2SO_4$ and amorphous water-ice abundances (Figure 15). In this case, it is likely that $MgCl_2$-abundant material upwelling from below the surface is reducing the signal of radiolytically produced $H_2SO_4$ and amorphous water-ice that was present on the surface, a hypothesis that confirms previous findings that magnesium-bearing species trace geological features and evidence for cryovolcanic activity on the surface (Carlson et al. 2005; Shirley et al. 2010; Ligier et al. 2016). However, we did not identify a correlation between $MgSO_4 \cdot 6H_2O$ and $H_2SO_4$, which was predicted by Brown & Hand (2013) and Vu et al. (2016) to exist owing to the formation of $MgSO_4 \cdot 6H_2O$ by radiolytically altered $MgCl_2$. This indicates that the upwelled material at Adonis Linea is relatively young and has not undergone significant radiolytic processing.

Enhanced abundances of NaCl and KCl did not correlate with the locations of lineae, suggesting that if transportation of material is occurring between Europa's subsurface ocean and its surface, its subsurface ocean may be more rich in magnesium, and more deficient in sodium and potassium, than previously suggested (Brown & Hand 2013; Trumbo et al. 2019). Further analyses of the distribution of brine composition across lineae are required in order to elucidate the composition of the subsurface ocean.

Finally, we demonstrated that the spectral mixture analysis and longitudinal abundance assessment used in this study are powerful techniques for determining which surface features are present in regions where the resolution of the imagery is too low to discern geological features; this can therefore inform region-of-interest observation selections for the Europa Clipper and proposed Europa Lander missions.

This study was funded by NASA under grant 80NSSC 17K0408 issued through the NASA Education Minority University Research Education Project (MUREP) as a NASA Harriett G. Jenkins Graduate Fellowship through the Aeronautics Scholarship & Advanced STEM Training and Research (AS&ASTAR) Fellowships. Galileo NIMS data were provided by the PDS Cartography and Imaging Sciences Node. Special thanks to Jesse Mapel (USGS) for assistance with debugging NIMS map projection issues and two anonymous reviewers for helpful comments that improved the quality of this manuscript.

## ORCID iDs

Jodi R. Berdis  https://orcid.org/0000-0002-8572-0036
James R. Murphy  https://orcid.org/0000-0003-1280-5167
Nancy J. Chanover  https://orcid.org/0000-0002-9984-4670